\documentstyle[aps,prl]{revtex}  
\begin{document}
\draft

\twocolumn[\hsize\textwidth\columnwidth\hsize  

\csname @twocolumnfalse\endcsname              

\title{\bf  $d \pi \pi$ decay of the $d^*$ dibaryon } 

\author { Chun Wa Wong}

\address{
Department of Physics and Astronomy, University of California, 
Los Angeles, CA 90095-1547}

\date{June 29, 1999}

\maketitle

\begin{abstract}

The $d^* \rightarrow d \pi \pi$ partial decay width has been 
calculated in a wave-function model for $d^*$ and $d$. It is 
found to be smaller than a previous estimate by a factor of 7. 
A previously proposed dependence on dibaryon sizes is confirmed. 
The large reduction found is caused partly by a change in 
the $d^*$ size used and partly by the need to match 
momenta in pion emissions, a feature not included in the 
previous estimate. 

\end{abstract}

\pacs{  PACS number(s): 14.20.Pt, 25.40.Ep }

     ]  

\narrowtext

Years ago, Goldman {\it et al.} \cite{Gold89} (referred 
to below as GMSSW) have estimated the $d \pi \pi$ partial decay 
width of 
the dibaryon $d^* (J^\pi = 3^+, T = 0)$ to be about 20 eV at 
the dibaryon mass $m^*$ 100 MeV above the $d \pi \pi$ threshold of 
about 2150 MeV/$c^2$. Though small, this width is of some 
interest as the total pion production cross section of $d^*$ 
from deuteron can be related to it. The result estimated by GMSSW
is 0.1 $\mu$b at an incident pion momentum of 580 MeV/$c$.

To estimate the decay width, GMSSW use an effective 
$dd^*\pi\pi$ vertex of the form  $(g/m_\pi^2)d_{\{\mu\nu\lambda\}}^*
d^\lambda \partial^\mu \pi_1 \partial^\nu \pi_2$, where  
$\{\mu\nu\lambda\}$ are symmetrized indices. The effective coupling constant $g$ can be related to the effective coupling constant 
$g_{\pi N \Delta}$ for 
$\Delta \rightarrow N\pi$ decay by the expression

\begin{eqnarray}
g \approx g_{\pi N \Delta}^2 (m_\pi/M_N)^2 f(d^* \rightarrow \Delta\Delta)
f(d \rightarrow NN)~,
\end{eqnarray}
where the vertex overlap factors are taken to be 
$f(d \rightarrow NN) \sim 1$ and 
$ f(d^* \rightarrow \Delta\Delta) \sim (r^*/r_d)^3 \sim 2^{-3}$,
where $r_d$ ($r^*$) is the deuteron ($d^*$) radius. In this 
approximate treatment, one of the vertex factors can always be set 
to one, since it is only the volume ratio that matters. The reduction 
shown for the second vertex factor represents the estimated 
difficulty of finding the same three quarks in each baryon in 
the smaller size in $d^*$ when the baryon was originally in the 
larger bound state that was $d$. This GMSSW approach to the 
$d \pi \pi$ partial width is elegant, but it has an Achilles 
heel, namely the inability to improve on the given choice of the 
vertex factor. It is therefore of some interest to determine if the reported 
estimate is reliable.

Recently, I have estimated the $NN$ and $\pi NN$ partial widths 
of $d^*$ using a perturbative method based on a baryon wave 
function for $d^*$ \cite{Won98,Won98a}. The purpose of this 
brief report is to illustrate how this more systematic approach can be 
applied to the calculation of the $d\pi\pi$ partial width. The calculated 
width shows the size effect proposed by GMSSW, but depends additionally
and strongly on the momentum matching required in the pion emissions, 
a feature that is completely missing in the GMSSW approach. 

For an S-wave deuteron state described by a realistic Bonn C wave 
function \cite{Mac87} and a three-term Gaussian approximation 
to a two-center baryon wave function for $d^*$, the width 
calculated here is about 3 eV at $m^* = 2250$ MeV/$c^2$. 
This estimate is smaller than that reported by GMSSW by a factor 
of 7. About half of this reduction is caused by our use of a 
smaller volume ratio, while the remaining half of the 
discrepancy comes from the effect of momentum matching among the 
wave functions required in the pion emissions.

The $d^* \rightarrow d \pi \pi$ decay in the lowest-order 
perturbation theory can be visualized as the decay of the two 
off-shell $\Delta$'s shown in Fig. \ref {fig-diagrams} in 
time-ordered diagrams. The resulting decay width is, according to 
the Fermi golden rule \cite{PDG96}:

\begin{eqnarray}
\Gamma & = & {1\over (2\pi)^5} \int 
\langle \vert (V_1G_{12}V_2)_{\rm fi} \vert ^2 \rangle_{\rm spin}
\nonumber \\
&& \times E_1 E_2 E_3 {\rm d}E_1 {\rm d}E_2
{\rm d}^2\Omega_1 {\rm d}\phi_2~,
\label{Gamma}
\end{eqnarray}
where $V_i$ is a pion-emission vertex from a baryon, and $E_1$ 
and $E_2$ are the pion energies in the $d^*$ rest frame. $G_{12}$ 
is the sum of Green functions for the two diagrams labeled $a$ 
and $b$. To avoid possible zeros in the energy denominators 
$D_i, \, i=a,b$, where for example

\begin{equation}
D_a = m^* - E_1({\bf k}_1) - E_N(-{\bf p}-{\bf k}_1) - E_\Delta({\bf p})~, 
\label{da}
\end{equation}
I give $m^*$ a total width $\Gamma_{tot}$. Consequently

\begin{equation}
G_{12} = {D_a\over D_a^2 + (\Gamma_{tot}/2)^2} + 
{D_b\over D_b^2 + (\Gamma_{tot}/2)^2}~.
\label{G12}
\end{equation}

  The Green-function factor $G_{12}^2$ depends on the angles. 
However, if it is approximated by a suitable angle-averaged value 
$\langle G_{12}^2 \rangle$, and if the baryons in both initial and 
final dibaryons are in relative orbital S states, the angle 
integrations shown in Eq. (\ref {Gamma}) can be performed 
analytically to give 
 
\begin{eqnarray}
\Gamma \approx {2\over (2\pi)^3} & \int & 
\langle G_{12}^2 \rangle \langle \vert (V_1V_2)_{\rm fi} \vert ^2 \rangle_{\rm spin, angle} \nonumber \\
&& \times E_1 E_2 E_3 {\rm d}E_1 {\rm d}E_2~.
\label{Gamma2}
\end{eqnarray}

  Although the baryon-baryon relative wave functions of both 
$d^*$ and $d$ will eventually be expressed as sums of Gaussians, 
it is sufficient to show the result for the spin- and 
angle-averaged squared matrix element in the integrand for 
single Gaussians such as

\begin{eqnarray}
\psi_{d^*}({\bf p}) = (\beta^{*2}/\pi)^{3/4} e^{-p^2/2\beta^{*2}}
\label{Gaussian}
\end{eqnarray}
for $d^*$. The result obtained after the integration over the 
internal momentum {\bf p} shown in Fig. \ref {fig-diagrams} is 

\begin{eqnarray}
\langle \vert (V_1V_2)_{\rm fi} \vert ^2 && \rangle_{\rm spin, angle} = 
{A^2\over 54}\left ({2\beta \beta^*\over B^2} \right )^3 {k_1^2k_2^2\over E_1 E_2} \nonumber \\
&& \times e^{-\alpha (k_1^2 + k_2^2)}[j_0(-ix) - \frac{1}{5} j_2(-ix)]~,
\label{v1v2}
\end{eqnarray}
where

\begin{eqnarray}
A = \Gamma_\Delta \left ( {3\over 4\pi^2} \right ) {M_\Delta\over k^{*3}E_N^*} e^{\alpha_0k^{*2}}, \qquad \qquad \nonumber \\
B^2 = \beta^2 + \beta^{*2}, \quad \alpha = \alpha_0 + {1\over 4B^2}, 
\quad x = {k_1k_2\over 2B^2}~.
\label{ABx}
\end{eqnarray}
Here $\Gamma_\Delta = 120$ MeV is the $\Delta$ width, $k^* (E_N^*)$ 
is the momentum of the decay pion (recoiling nucleon) in the rest 
frame of the decaying $\Delta$, and 

\begin{eqnarray}
\alpha_0 = r_p^2/3 = 0.12\; {\rm fm}^2
\label{alpha0}
\end{eqnarray}
is the parameter, taken to be the same in N and $\Delta$, 
describing the Gaussian wave functions of quarks inside these 
baryons.  The baryon form factors in the pion-emission vertex give 
rise to the $\alpha_0$ term in $\alpha$.

  When the baryon wave function of the dibaryon $d^*$ is 
approximated by a single Gaussian, the parameters $\beta^*$ and 
the m.s. baryon momentum $<{\bf p}^2>$ inside the dibaryon are 
related to the dibaryon radius $r^*$ as

\begin{eqnarray}
\beta^* = \sqrt{3/8}/r^*~, \quad <{\bf p}^2> = {9\over 16 r^{*2}}~.
\label{beta*}
\end{eqnarray}
The average baryon energies to be used in the Green-function 
factor $<G_{12}^2>$ are

\begin{eqnarray}
E_\Delta ({\bf p}) \approx \sqrt{M_\Delta^2 + <{\bf p}^2>}~, 
\quad\quad \nonumber \\
E_N ({\bf p}+{\bf k}) \approx \sqrt{M_N^2 + <{\bf p}^2> + k^2}~.
\label{EDEN}
\end{eqnarray}

  Certain features in our results are worth pointing out: The 
$\beta$-dependent factor in Eq. (\ref {v1v2})  can be expressed approximately in terms of baryon radii as	

\begin{eqnarray}
 \left ({\beta \beta^*\over B^2} \right )^3  \approx \left (r^*\over r_d \right)^3.
\end{eqnarray}
This is the only feature included in the GMSSW vertex factor. 
Note however that if we use $r_d = 2.0$ fm and $r^* = 0.7$ fm, 
then this volume ratio is smaller than that used by GMSSW by a 
factor of 3. It is possible that GMSSW have used a larger 
volume ratio in the expectation that in the quark-delocalization 
model the quarks from a baryon constituent of the dibaryon are 
more spread out over the dibaryon than in a conventional 
baryon-baryon bound state. Unfortunately, this expectation has not 
been quantified in GMSSW. 

  A quantitative study of this effective size effect of 
delocalization on the $d\pi\pi$ decay width is not easy 
because delocalizaed quark wave functions \cite{Gold89,Wang92} 
are so complicated when antisymmetrization and angular momentum 
projections are made. In addition, delocalized short-distance wave functions will also have to be used in the deuteron for overall consistency. In the more literal approach taken here, this 
important question will be left as an open problem for future study. 
On the other hand, certain other features of the problem, such as an increase in the dibaryon size, can be studied rather easily in the simple model used in this paper.  

  The remaining features of Eq. (\ref {v1v2}) are not included in 
the GMSSW vertex factor. They describe the requirements of 
momentum matchings on pion emissions. Note that the parameter $B^2$ 
is dominated by $\beta^{*2}$, which is proportional to the m.s. 
baryon momentum contained in $d^*$.  For $r^* = 0.7$ fm, 
$B^2 = 0.29$ fm$^2$, which is larger than $\alpha_0$ by a factor 
2.4. This shows that the dependence of the parameter $\alpha$ on 
baryon sizes in the inelastic baryon form factors is 
significantly weaker than the dependence on $r^*$. If the $\Delta$ 
radius $r_\Delta$ is different from the nucleon radius $r_p$, the parameter $\alpha_0$ should be replaced by

\begin{eqnarray}
\alpha_0 \left (2r_\Delta^2\over r_p^2 + r_\Delta^2 \right )
\approx \alpha_0 \left ( r_\Delta\over r_p \right ).
\end{eqnarray}
Since $r_\Delta$ is expected to be larger than $r_p$ by only 10 
($\approx 5$) \% in the Massachusetts Institute of Technology 
bag model \cite{DeG75} (many potential models \cite{Won82}), 
the overall effect in the 
parameter $\alpha$ is only 3(2)\%, which is quite negligible. 
In contrast, results will be shown below where the parameter 
$\alpha$ changes by a factor of almost 2. I therefore conclude 
that the use of the same baryon size for both $\Delta$ and $N$ 
is justified both here and in the ``delocalization'' model of  
\cite{Gold89,Wang92}. Without this simplification of equal baryon 
sizes, the delocalization model would be even harder to execute technically.

  The total width $\Gamma_{tot}$ of $d^*$ is also needed in 
the calculation. It is known to increase with increasing 
$d^*$ mass $m^*$, rising from about 1 MeV at $m^* = 2100$ 
MeV/$c^2$ to about 10 MeV at $m^* = 2350$ 
MeV/$c^2$ \cite{Won98,Won98a}. However, the presence of 
$\Gamma_{tot}$ is unimportant at the low end of this mass range, 
because the energy denominators $D_i$ are large there. Beyond 
the $\Delta N\pi$ threshold at 2310 MeV, however, the 
energy denominators could become small; the choice of 
$\Gamma_{tot}$ then becomes important. For this reason, I use a 
nonzero width in the calculation, but choose for simplicity a 
single constant value of $\Gamma_{tot} = 10$ MeV appropriate to 
the high end of the mass 
range studied here. The effect of using smaller values of 
$\Gamma_{tot}$ will be explicitly shown below.

  The single S-state Gaussian wave function for the deuteron 
depends on the parameter $\beta = \sqrt{3/8}/r_d$, where 
$r_d  = 1.967$ fm is the radius of the deuteron wave function. 
The calculated 
$d\pi\pi$ partial width for $r^* = 0.7$ fm is shown in Fig. \ref 
{fig-widths} as a thin solid curve for a range of $m^*$. The value 
at $m^* = 2250$ MeV/$c^2$ is 2.0 eV, ten times smaller than the 
estimate given by GMSSW. Besides the decrease by a factor of 3 
already pointed out and discussed previously, there is a 
remaining reduction by another factor of 3 which should be 
attributed to the technical improvements made in the present calculation.

  For a more realistic S-state wave function, I use the three-term Gaussian fit to the Bonn C deuteron S-wave wave function 
obtained in \cite{Won98}. The fitted S-state probability of 
94.34\% (versus 94.39\% for the original Bonn C wave function) 
has been renormalized back to 100\% for a pure S state. The 
resulting $d\pi\pi$ partial width is shown in Fig. \ref {fig-widths} 
as a solid (dashed, long-dashed) curve for 
$r^* = 0.7 \;(0.9, 0.5)$ fm. 

  These curves show that the decay width for the realistic Bonn C 
wave function is larger than that for the single Gaussian wave 
function rather uniformly over the mass range. At $m^* = 2250$ MeV, 
the increase is by a factor of 1.9 at $r^* = 0.7$ fm. Since the 
deuteron radius is essentially the same in these calculations, 
the change must have come entirely from the increase in the 
high-momentum components in the Bonn C wave function.

  Figure \ref {fig-widths} also shows results calculated with the 
same Bonn C wave function but different $d^*$ radii. In the simple volume scaling model of GMSSW, the decay width for $r^* = $ 0.7 fm 
would be increased (decreased) a factor of 2.2 (2.7) when $r_d^*$ 
is increased to 0.9 fm (decreased to 0.5 fm). The actual calculated 
factor at $m^* = 2250$ MeV turns out to be 1.6 (2.7) for the Bonn C deuteron, and 1.9 (3.3) for the single-Gaussian deuteron. Thus the overall volume scaling effect proposed by GMSSW appears to be present, 
but in a more complicated form that also depends on other details of 
the wave function. The main reason for the complication is 
that the parameter $\alpha$ also changes drastically with $r^*$ --- 
from 0.40 fm$^2$ for $r^* = 0.7$ fm to 0.28 (0.58) fm$^2$ for 
$r^* = 0.5$ (0.9) fm.

  The additional sensitivity to the high-momentum components of 
the deuteron wave function shown in the present calculation 
suggests that the deuteron D-state component, 
though constituting only 5\% of the deuteron, might have a 
disproportionate effect on these decay widths. Its inclusion is 
easy to visualize, but tedious to execute. I believe that its 
inclusion will not change the order of magnitude of the partial 
widths estimated here --- after all, the effect of changing the 
S-state wave function from the smooth single Gaussian to the 
realistic Bonn C wave function is an increase by ``only'' a 
factor of two.

  I now turn to the effect of using smaller total decay widths
in the calculation. The results for $\Gamma_{tot} = 5$ (2.5) MeV 
are shown in Fig. \ref {fig-totalw} as a dashed (dotted) curve, 
and compared with the corresponding result for 
$\Gamma_{tot} = 10$ MeV from Fig. \ref {fig-widths}, reproduced 
here as a solid curve. The calculated width can be seen to develop significant dependence on the total decay width only well above 
the $\Delta N \pi$ threshold of 2310 MeV.

  The single Gaussian wave function for $d^*$ is of course a 
rather crude approximation to the model of ``delocalized'' 
quarks in $d^*$ \cite{Gold89,Wang92}. 
The least improvement one could make is to use a two-center 
Gaussian wave function for each {\it baryon} proportional to 
exp[-({\bf x} - {\bf s})$^2$/2] + exp[-({\bf x} + {\bf s})$^2$/2],
where {\bf x} is the (dimensionless) relative baryon-baryon 
coordinate and 2{\bf s} is the separation between the 
two centers. The projected S-state component of this relative wave function has the form

\begin{eqnarray}
\psi_0(x,s) = N_0 {1\over x} \left [ e^{-(x-s)^2/2} - 
e^{-(x+s)^2/2} \right ]~.
\label{projWF}
\end{eqnarray}
The lowest-order effect of $s$ can be included in the 
single Gaussian wave function by matching the m.s. radius 

\begin{eqnarray}
\langle x^2 \rangle = {1\over 2} + {s^2\over 1-e^{-s^2}} 
\approx {3\over 2} + {s^2\over 2}~. 
\label{series}
\end{eqnarray}
For small $s$, one sees the separate contributions from 
the zero-point motion of the baryons in the two harmonic 
oscillator potentials and from the separation $2s$ of these 
potentials. If one starts with quark wave functions described by a 
size constant $b = 0.6$ fm, the relative baryon-baryon motion for 
the six-quark state will be described by the size constant 
$b_r = b/\sqrt{6}$. Hence $r^* = b_r \: \langle x^2 \rangle^{1/2}$. 
For a potential separation $2S = 1.40$ fm calculated for the $d^*$ 
\cite{Gold89,Wang92}, $s = S/b_r = 2.86$, and hence $r^* = 0.72$ 
fm. This is close to the middle of the three values used in Fig. 
\ref {fig-widths}.

  It is easy to go beyond this rough approximation and actually fit 
the momentum wave function of the projected two-center Gaussian, 
namely $\sim {\rm exp}[-(b_rp)^2/2] j_0(pS)$, as a sum of 
three Gaussians

\begin{equation}
\psi_{\rm 2c}(p) \approx  \sum_{i=1}^3 c_i \psi_i(p),
\end{equation}
where $\psi_i(p)$ is a normalized Gaussian wave function. In order 
to emphasize the stronger high-momentum components introduced by 
the oscillating factor $j_0(pS)$, the range parameters 
are obtained by minimizing the {\it percentage} m.s. deviation. 
The fitted result is

\begin{equation}
\mbox{\boldmath $\gamma$} =  (\gamma_1, \gamma_2, \gamma_3) = 
0.9961\, (1, 1.25, 1.46)/b_r^2~,
\end{equation}
where $\gamma_i = 2\beta_i^2$. The expansion coefficients, 
renormalized from the fitted normalization of 1.0019 back to 1, are

\begin{equation}
{\bf c} = (c_1, c_2, c_3) = (14.8311, -27.4124, 13.3505)~.
\end{equation}
The $d\pi\pi$ partial decay width can now be calculated with 
this improved $d^*$ wave function but with everything else 
treated in the same way as the single-Gaussian case with 
$r^* = 0.7$ fm. The results, shown as solid circles in Fig. \ref 
{fig-widths}, are about 20\% smaller than those for the single 
Gaussian $d^*$ wave function. The radius of the fitted $d^*$ wave 
function, at 0.72 fm, is actually marginally larger, but the 
decrease in the calculated decay width shows that it is the 
additional high-momentum components in the $d^*$ wave function 
that dominate the result. 

  Although the present calculation is a significant improvement 
over that of GMSSW, it is also not a quantitative calculation. 
Too many approximations have to be made to reduce the problem into 
manageable form. The most important features that should be 
included in a realistic calculation include the following: 
(1) higher-order Feynman diagrams, especially those 
involving intermediate- and final-state interactions; (2) quark 
antisymmetrization between baryons and quark delocalization 
effects in both $d^*$ and $d$; (3) deuteron D state and perhaps 
better short-distance wave functions for both $d$ and $d^*$, in 
both $NN$ and exotic channels; (4) improved treatment of the 
energy denominator, requiring the use of full angle integrations; 
(5) better treatment of the $\Delta N \pi$ vertex; (6) a better 
choice of the full decay width, but only when the dibaryon mass 
is above the $\Delta N \pi$ threshold. It is clear, however, 
that these improvements will require much more extensive 
calculations than those undertaken here.

\acknowledgements

I would like to thank Terry Goldman for urging me to improve on the single-Gaussian approximation for $d^*$ and Fan Wang for providing 
the parameters of the two-center Gaussian wave function.

\centerline{\bf Figure Captions}

\begin{figure}
\caption{ The two leading-order time-ordered diagrams for the decay
$d^* \rightarrow d\pi\pi$. }
\label{fig-diagrams}
\end{figure}

\begin{figure}
\caption{ The $d^* \rightarrow d\pi\pi$ partial decay width 
$\Gamma_{d\pi\pi}$ as a function of the $d^*$ mass $m^*$ for 
single-Gaussian and Bonn C deuteron S-state wave functions, and 
for single-Gaussian and two-center $d^*$ wave functions. The number 
shown in the legend is the $r^*$ radius in fm. }
\label{fig-widths}
\end{figure}

\begin{figure}
\caption{ The $d^* \rightarrow d\pi\pi$ partial decay width 
$\Gamma_{d\pi\pi}$ as a function of the $d^*$ mass $m^*$ for 
Bonn C deuteron S-state wave function and single-Gaussian $d^*$ 
wave function (with $r^* = 0.7$ fm) when the input total width 
of $d^*$ is varied. }
\label{fig-totalw}
\end{figure}

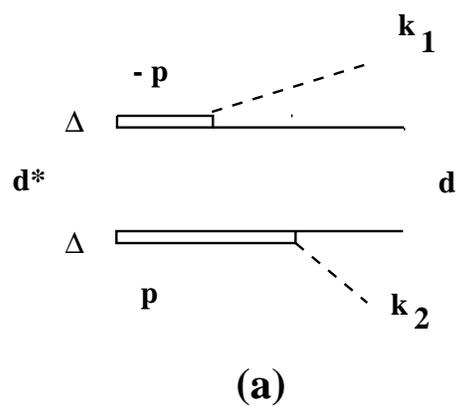
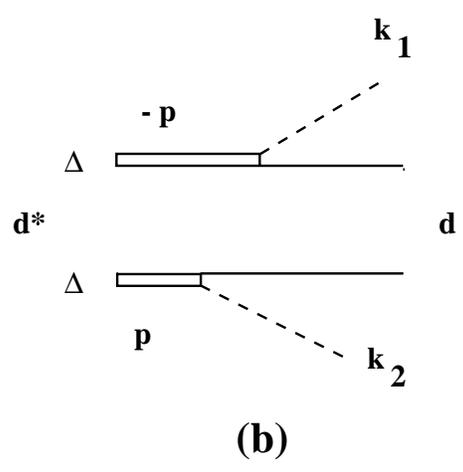
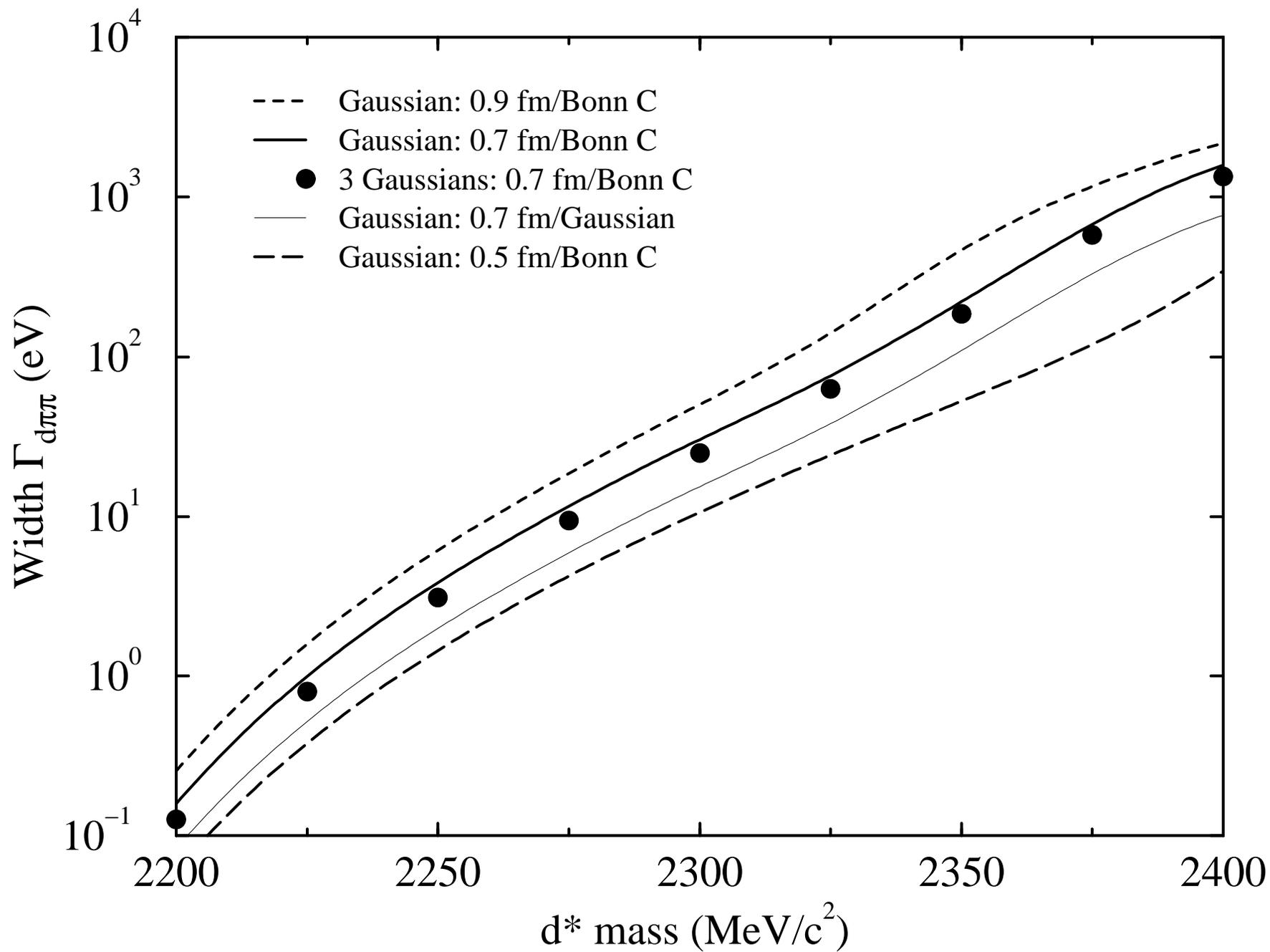
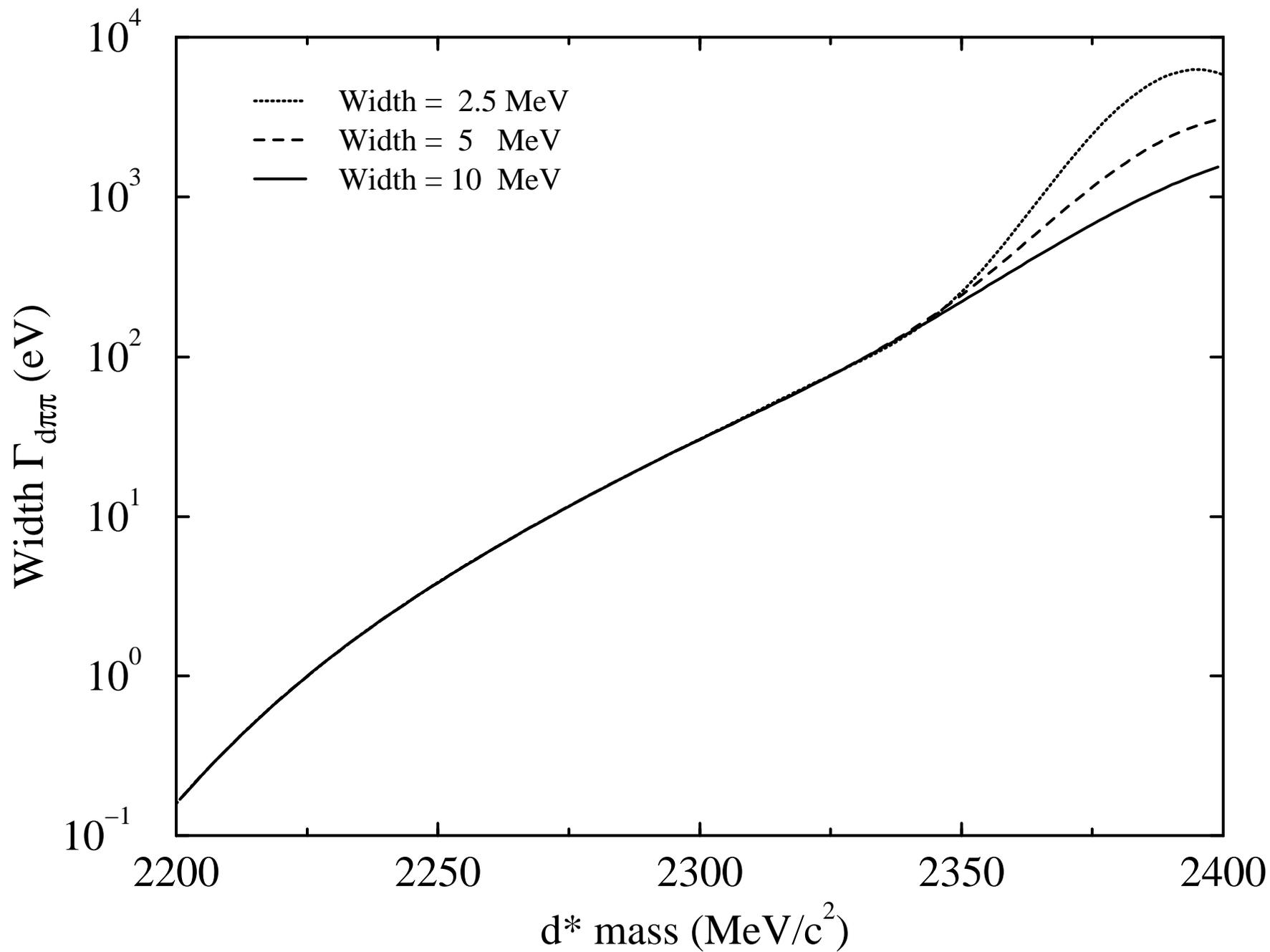
\end{document}